\documentclass{iaus}
\usepackage{graphicx}

\title[MASH-MEN Project] 
{Planetary nebulae and their mimics:\\ the MASH-MEN Project }

\author[Rozenn Boissay, Quentin A. Parker, David J. Frew \& Ivan Bojicic]   
{Rozenn Boissay$^{1,2,3}$, Quentin A. Parker$^{2,3,4}$, David J. Frew$^{2,3}$\\ \and Ivan Bojicic$^{2,3}$
}

\affiliation{$^1$Observatoire Astronomique, Universit\'{e} Louis Pasteur, 67000 Strasbourg, France \\ email: {\tt rozenn.boissay@gmail.com} \\[\affilskip]
$^2$Department of Physics, Macquarie University, North Ryde, NSW 2109, Australia  \\email: {\tt quentin.parker@mq.edu.au} \\[\affilskip]
$^3$Macquarie University Research Centre in Astronomy, Astrophysics \& Astrophotonics\\[\affilskip]
$^4$Australian Astronomical Observatory (AAO), Epping, NSW 1710, Australia }

\pubyear{2011}
\volume{283}  
\pagerange{119--126}  
\jname{Planetary Nebulae: an Eye to the Future}
\editors{Arturo Manchado, Letizia Stanghellini \& Detlef Sch\"onberner, eds.} 
\begin{document}

\maketitle

\begin{abstract}
The total number of true, likely and possible planetary nebulae (PN) now known in the Milky Way is about 3000, approximately twice the number known a decade ago. The new discoveries are a legacy of the recent availability of wide-field, narrowband imaging surveys, primarily in the light of H$\alpha$. The two most important are the AAO/UKST SuperCOSMOS H$\alpha$ survey – SHS and the Isaac Newton photometric H$\alpha$ survey  - IPHAS,  which are responsible for most of the new discoveries.  A serious problem with previous PN catalogues is that several different kinds of astrophysical objects are able to mimic PN in some of their observed properties leading to significant contamination. These objects include H~II regions and Str\"{o}mgren zones around young O/B stars, reflection nebulae, Wolf-Rayet ejecta, supernova remnants, Herbig-Haro objects, young stellar objects, B[e] stars, symbiotic stars and outflows, late-type stars, cataclysmic variables, low redshift emission-line galaxies, and even image/detector flaws. PN catalogues such as the Macquarie/AAO/Strasbourg H$\alpha$ Planetary Nebula catalogue (MASH) have been carefully vetted to remove these mimics using the wealth of new wide-field multi-wavelength data and our 100\% follow-up spectroscopy to produce a compilation of new PN discoveries of high purity. 
During this process significant numbers of PN mimics have been identified. 
The aim of this project is to compile these MASH rejects into a catalogue of Miscellaneous Emission Nebulae (MEN) and to highlight the most unusual and interesting examples. A new global analysis of these MEN objects is underway before publishing the MEN catalogue online categorizing objects by type together with their spectra and multi-wavelength images. 
\end{abstract}

\firstsection 
\section{Introduction}

The MASH Miscellaneous Emission Nebulae catalogue (MASH-MEN) represents the compilation of about 450 new emission line sources identified as PN candidates but now removed as contaminants mostly prior to publication of MASH-I (\cite[Parker \etal ~2006]{ParkerAl2006}) and MASH-II (\cite[Miszalski \etal ~2008]{MiszalskiAl2008}). This was achieved by careful application of spectral and multi-wavelength image diagnostic criteria as described in \cite[Frew \& Parker (2010)]{FrewParker2010}.

\section{Methodology}

Several kinds of astrophysical objects mimic planetary nebulae and pre-MASH PN catalogues still contain significant numbers of contaminants that are being detected and eliminated via our newly developed methodology. To classify MEN objects into categories, we first collect as much information as possible, i.e. multi-wavelength images and spectra (from surveys as SHS, SSS, DSS, IPHAS, 2MASS, MSX, MIPS, GLIMPSE, and WISE). The H$\alpha$/SR quotient image is effective for revealling the extent of any emission to emphasize the MEN objects. It is also useful to create RGB images whiwh can reveal a reflection component from the B-band. When all these data are available, we are able to classify objects into categories thanks to their morphological and spectroscopic criteria where the overall body of evidence is used in combination with a decision tree we created. Other tools such as diagnostic plots using nebular emission-line ratios are also used to distinguish PNe, Herbig-Haro objects, SNRs, and HII regions. Photometric data are useful to identify symbiotic stars or close-binary CSPN. Other properties used for the identification of the MEN objects are detailed in \cite[Frew \& Parker (2010)]{FrewParker2010}.


\section{Examples and results}

\begin{figure}[h!]
\begin{center}
 \includegraphics[scale=0.42]{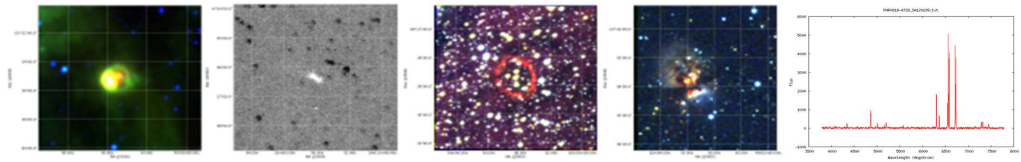} 
 \caption{From left to right: PHR0703-1035: HII region (WISE image); PHR1811-0436: symbiotic star with outflows (quotient image); PHR1633-4928: Pop.I WR star with ejecta (SHS image); PHR0901-4743: stars forming region (2MASS image); PHR0818-4728: SNR (spectrum).}
   \label{fig1}
\end{center}
\end{figure}

All 450 MEN sources (see examples in figure \ref{fig1}) have been classified into about a dozen different object categories. Unsurprisingly, HII regions (39\% of MEN objects) and symbiotic stars (22\%) are most often confused with PNe. The MEN catalog contains also 8\% SNRs, 5\% flaws, 4\% galaxies, 4\% reflection nebulae, 3\% YSOs and Herbig-Haro objects, 3\% cataclysmic variables, 2\% clusters, 2\% B[e] stars, 1\% Wolf-Rayet ejecta, and under 1\% LBVs. Finally, as better data became available, several bona-fide PN previously removed from MASH can now be re-integrated as a result of this project uncovering their true PN nature (6\% of the MEN objects).

\section{Conclusion}

The MASH-MEN Project was instigated first to house the many interesting but non-PN mimics uncovered during the MASH survey and subsequently to classify the mimics according to type into a new catalogue of miscellaneous emission nebulae (MEN). Application of our careful diagnostic processes have resulted in the MASH PN catalogues being of high purity. Some of these contaminants are very interesting in their own right and deserve follow-up and further study. We are now applying the same processes to other extant PN catalogues to remove their many mimics.

\end{document}